\documentclass[%
reprint,                     
amsmath,amssymb,
aps,
]{revtex4-2}

\usepackage{graphicx}
\usepackage{dcolumn}
\usepackage{bm}

\usepackage{times}
\usepackage{hyperref}
\usepackage{ulem}
\usepackage{comment}

\usepackage{xcolor}

\begin{document}
	
	\preprint{APS/123-QED}
	
	\title{ 
	Bell Nonlocality Test on Two-Mode Squeezed Output Generated in Double-Cavity Optomechanical Systems
}

\author{Souvik Agasti}
\email{souvik.agasti@uhasselt.be}

\affiliation{
	IMOMEC division, IMEC, Wetenschapspark 1, B-3590 Diepenbeek, Belgium
}%
\affiliation{
	Institute for Materials Research (IMO), Hasselt University,	Wetenschapspark 1, B-3590 Diepenbeek, Belgium
}%


\begin{abstract}
	
	We explore here how to generate a two-mode squeezed output using reservoir engineering in a double-cavity optomechanical system coupled to a common mechanical resonator. 
	Such hybrid platforms are experimentally accessible in electro-optomechanical interfaces and are relevant for high-fidelity state transfer {by means of quantum teleportation} and metrological applications. By examining violations of the CHSH Bell inequality, we demonstrate that maximal squeezing does not necessarily imply nonlocality; instead, nonlocal correlations can emerge in states with lower squeezing. Furthermore, by analyzing the CHSH inequality across different cavity finesse values, we find that the parameter region supporting nonlocality can broaden even as the squeezing region shrinks. Across all regimes considered, 
	our results 
	emphasize the crucial influence of the mixedness of the state in determining the inverse relationship between squeezing and nonlocality.

\end{abstract}

\maketitle


\section{Introduction}\label{Introduction}

Continuous-variable (CV) entangled Gaussian states have attracted considerable attention due to their applications in precision measurement \cite{precision_measurement1,precision_measurement2, Squeezing_mechanics_Clerk}, quantum communication \cite{quantum_communication1, quantum_communication2, quantum_communication3}, and high-fidelity state transfer \cite{quantum_teleportation, Clerk_Fidelity, vitali_TMS_generator}. The most popular example of such a bipartite Gaussian CV state is the two-mode squeezed vacuum (TMSV) state. The entanglement shared between its constituent modes is of fundamental and practical interest, with applications ranging from quantum memories \cite{Quantum_memory} to gravitational-wave metrology \cite{study4roadmap, my_BAE}. {E}ntanglement motivates the test of a hidden variable in such systems that puts a limitation on local realism, which has been determined through Bell’s inequality \cite{mypaper_TMSV_filter, mypaper_TMSV_filter_STD}. Quantum nonlocality for spatially separated systems was formalized by Clauser, Horne, Shimony, and Holt (CHSH) \cite{bell_CHSH, bell_CH}, and later reformulated in phase space by Banaszek and W\'odkiewicz using Wigner quasiprobability distributions \cite{Banaszek_TMSV_bell, Banaszek_TMSV_bell_PRL}. These approaches provide criteria for nonlocality in bipartite Gaussian CV states and highlight the important role of state mixedness \cite{agasti_nonlocally_CV}, consistent with the fact that \textit{all nonlocally realizable bipartite Gaussian states must be entangled}.

Squeezed light is most commonly generated via parametric down-conversion in Kerr nonlinear media or through four-wave mixing in atomic vapors, with substantial experimental progress in both directions \cite{squeeze_light_sourse, TMS_generation, Squeezing_generation}. However, Kerr-based schemes typically involve non-Gaussian dynamics, leading to phase-space distortions that limit the achievable squeezing \cite{Agasti_Kerr_PhyScr, Agasti_Kerr_JOSAB}. Even at strong nonlinearities, phase-space twisting and non-Gaussian features become increasingly pronounced, and optimal squeezing is attained only over a restricted interaction-time window. Nevertheless, nonlinear processes tend to amplify input noise, thereby degrading the quality of squeezing.


As an alternative, optomechanical systems can exhibit an effective Kerr-type nonlinearity \cite{Equivalence_optomechanics_Kerr}, motivating their use as platforms for generating nonclassical states. In particular, double-cavity optomechanical systems with reservoir engineering have emerged as promising candidates for generating two-mode squeezing (TMS) \cite{Clerk_Fidelity, Reservoir_Engineering_two_mode, vitali_TMS_generator}. Such architectures can be realized, for instance, by coupling optical cavities to microwave superconducting circuits \cite{Teufel_om_experiment}, or by incorporating high-Q membranes within optical cavity setups \cite{double_cav_alternate}. Systems of this kind--featuring two cavities coupled to a common mechanical resonator--have also been employed in back-action-evading measurements relevant to gravitational-wave detection \cite{my_BAE} and in precision force sensing \cite{force_sensing_taylor}.

In this work, we employ reservoir engineering in a double-cavity optomechanical system to generate strongly entangled output states that serve as sources of TMS. Our approach combines sideband cooling in one cavity--effectively enabling state exchange between optical and mechanical modes--with amplification in the other cavity, which drives the generation of entanglement. Furthermore, we {characterize} the TMS state by exploiting {correlations} between them, {highlighting non-monotonicity between entanglement and nonlocality.
Recent advances in controlling coupled quantum optical and mechanical degrees of freedom have established cavity optomechanical systems as vital building blocks for quantum information and communication technologies. Beyond applications like gravitational wave detection, these platforms offer unique opportunities to test fundamental physics, from probing macroscopic quantum entanglement to searching for quantum gravitational phenomena. In this spirit, we investigate here the CHSH test of Bell's measurement in an optomechanical system.}

Given that optomechanically generated TMS states are Gaussian and highly entangled, which has been studied previously \cite{Reservoir_Engineering_two_mode, vitali_TMS_generator}, it is natural to investigate their Bell nonlocality. {Importantly}, the {discrepancy} between entanglement and non-locality in the bipartite measurement in optomechanical systems has not been understood properly, {especially} the roles played by the purity of the state. Beyond foundational interest, such tests are relevant for certifying security in quantum cryptographic protocols \cite{cryptographic_Bell}. 	A few ideas for testing the violation of CHSH inequalities in an electro- and optomechanical setting have been proposed before \cite{Proposal_Bell_test_Electromechanics, Proposal_Bell_test_optomechanics, Optomechanical_Bell_Test}; however, they are based on different configurations than the one discussed here. Here, we instead focus on Bell tests applied to specific spectral components of the output Bogoliubov modes, selected via filtering. This enables us to directly probe the relationship between entanglement (or TMS) and nonlocality, with particular emphasis on the role of mixedness of the state. In general, even though it is considered that entanglement plays {a} fundamental role for nonlocal quantum measurements, nonlocality and entanglement are not only different properties but are found to be inversely related in two-qubit systems \cite{inverse_nonlocality_entanglement}. {Although entanglement and nonlocality are distinct physical phenomena, the latter has frequently been conflated with evidence of the former \cite{ent_nonL_rev}. Indeed, all non-product states with classical correlations can violate Bell’s inequalities \cite{Werner_classical_bell, Gisin_Bell, Doherty_bell}. It has been proven that entangled mixed states do not always violate a Bell's measurement \cite{Barrett_bell}. However, strong quantum nonlocality can manifest independently without entanglement \cite{Quantum_Nonlocality_without_Entanglement}; consequently, only a restricted subset of nonlocally realizable states can serve as witnesses for quantum entanglement \cite{Buscemi_Nonlocality_Entanglement}. }
Therefore, it also becomes interesting to investigate if they show any {non-monotonic} relation in case of optomechanical systems, and overall, bipartite Gaussian CV systems.

The article is organized as follows: in the next section, we show how we can generate TMS Bogoliubov modes in the output of a double-cavity optomechanical system using a blue and a red detuned pump. Furthermore, we exploit quantum correlation between the specific spectral components of their outputs to study the discrepancy between TMS and the violation of the CHSH inequality. By varying the finesses of the cavities, we show that the parameter regime supporting nonlocality can broaden even as the region exhibiting TMS shrinks. The article {concludes by} investigating the role of the quality-factor of the mechanical oscillator and the temperature of its reservoir on the TMS and nonlocality between outputs, emphasizing throughout the critical role of state mixedness in governing their interplay. {This work is particularly significant in two key aspects: first, it presents a technique for generating relatively pure TMS outputs, which is of substantial interest to the research community. Second, it reveals a unique non-monotonic behavior between entanglement and nonlocality, observed here for the first time in CV Gaussian states. Furthermore, the article introduces a methodology for performing Bell measurements on Gaussian states, particularly within optomechanics and evaluating CHSH violations as a measure of nonlocality.}


\begin{figure*}
	\includegraphics[width= 0.45 \linewidth]{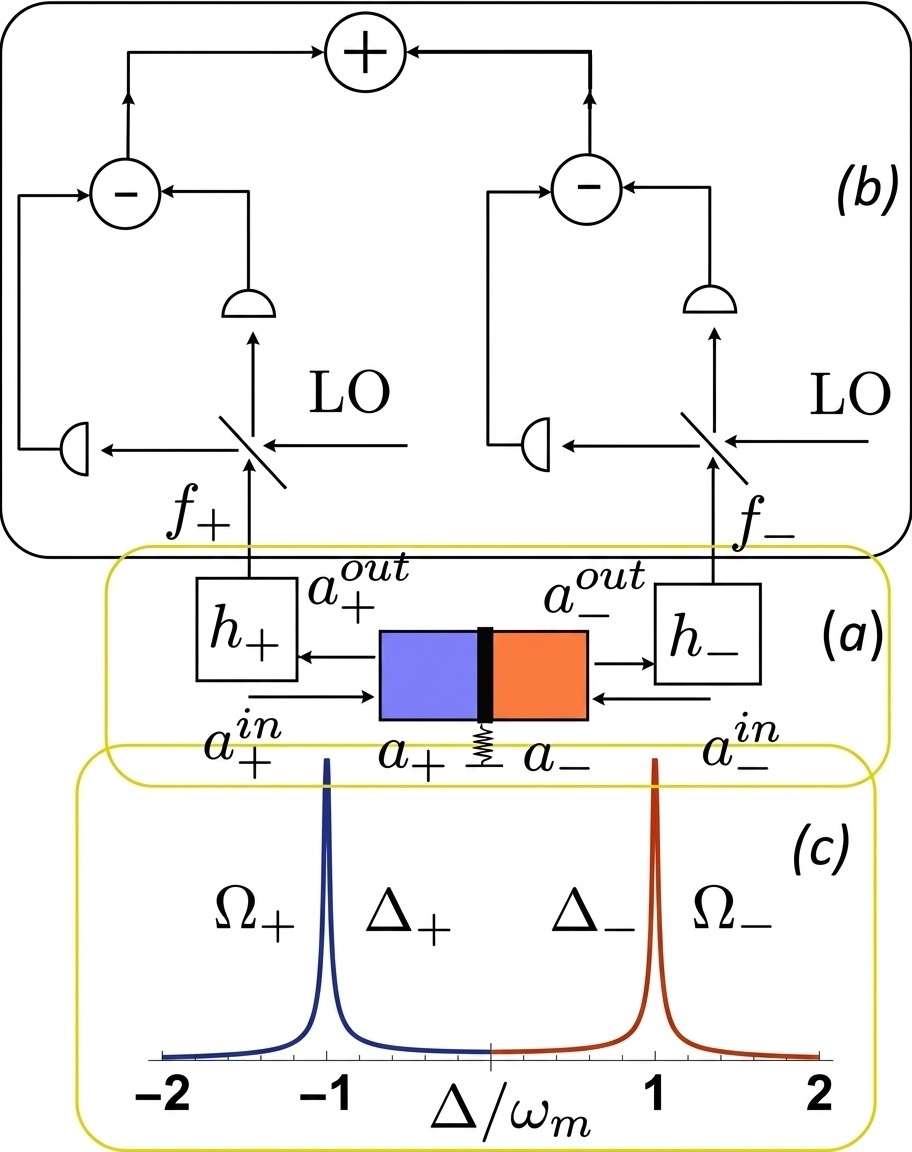}
	\caption{  Block diagram of (a) an optomechanical system where two cavities \textbf{are} coupled to a common mechanical oscillator and the filtering of modes at output. (b) Homo/heterodyne measurement of filtered output and the bipartite quantum correlation (LO refers to the local oscillator). (c) {Illustration of the ranges of frequencies at which the red (detuning frequency $\Delta_-$ = mechanical oscillation frequency $\omega_{m}$) and blue detuned ($\Delta_+ = - \omega_{m}$) cavities resonate. The filters operate on the frequencies around their corresponding cavity detunings, i.e. central frequencies of the filters are $\Omega_- \approx \Delta_-$ and $\Omega_+ \approx \Delta_+$.   }
	}\label{Block_diagram}
\end{figure*}

\begin{figure*}
	\includegraphics[width= 1 \linewidth]{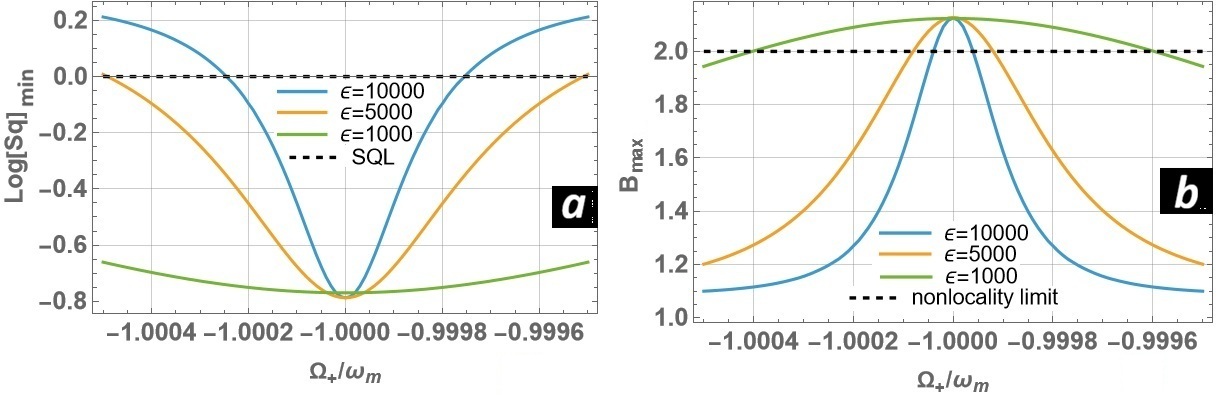}
	\caption{ (a) TMS ($\log[S_q]_{\min}$) and (b) maximal value of Bell function ($B_{\max}$) vs normalized output filter frequency for different bandwidths of filters $\epsilon = \tau \times \omega_m$. {The other} fixed central frequency of the filter at red detuned port {is} $\Omega_- = \omega_m$. The cavity detuning frequencies are $-\Delta_+ = \Delta_- = \omega_{m} = 2\pi \times 10 $ MHz. Other parameters are, for example cavity {linewidth} $\kappa_+ = \kappa_- = \kappa =0.02 \, \omega_m$, effective optomechanical couplings
		$ G_- = 0.15 \,\omega_m , G_+ = 0.2 \,G_- $, the quality factor {of the mechanical oscillator} $Q \equiv \omega_m/\gamma_m = 1.5 \times 10^5,$ and the temperature is around $25 mK$ which has corresponding thermal population {of the mechanical reservoir} $
		n_m \approxeq 500$. 
		This parameter set remains consistent with the parameters used in experiments conducted in
		\cite{Teufel_om_experiment}, which has also been followed in \cite{vitali_TMS_generator}. The only exception considered here is a lower mechanical quality factor than  \cite{Teufel_om_experiment}, so that {a} mirror can have a highly reflective coating.
	}\label{SQZFilter_nonloc}
\end{figure*}

\section{Model: Double-Cavity Optomechanical System}

\subsection{The Hamiltonian}

The Hamiltonian of a double-cavity optomechanical system coupled to a common mechanical oscillator is 

\begin{equation} \label{system_Hamiltonian}
	H_{OM} = \omega_m b^\dagger b + \sum_{k=\pm} \omega_{k}\, a_k^\dagger a_k + g_k \, a_k^\dagger a_k (b^\dagger + b),
\end{equation}
where $a_k$ ($a_k^\dagger$) denote the annihilation (creation) operators of the cavity modes with $k \in [+,-]$, and $b$ ($b^\dagger$) corresponds to the mechanical mode. The resonance frequencies of the cavity and mechanical modes are $\omega_k$ and $\omega_m$, respectively, with dissipation rates $\kappa_k$ and $\gamma_m$. The single-photon optomechanical coupling strengths are given by $g_k = (\omega_k/L_k)\sqrt{\hbar/(m\omega_m)}$, where $L_k$ is the effective cavity length and $m$ is the effective mass of the mechanical resonator. In the regime where the mechanical frequency is much smaller than the cavity free spectral range ($\omega_m \ll \mathrm{FSR} \sim c/L_k$), each cavity can be treated as single-mode, as the external drives predominantly address a single resonance. The cavities are driven by lasers with frequencies $\omega_k^L$, yielding the driving Hamiltonian
\begin{equation}
	H_{\mathrm{drive}} = \sum_{k=\pm} \left(a_k^\dagger E_k e^{-i\omega_k^L t} + \mathrm{h.c.}\right),
\end{equation}
where the drive amplitudes are $|E_k| = \sqrt{2P_k \kappa_k/(\hbar \omega_k^L)}$ and depend on the input powers $P_k$.

Entanglement between the two cavity modes can be generated by driving one cavity (labeled $+$) with a blue-detuned pump and the other ($-$) with a red-detuned pump. The corresponding detunings are defined as $\Delta_k = \omega_k - \omega_k^L$. 
The basic block diagram is presented in Fig. \ref{Block_diagram}, where two different drives of the cavities fix the cavity detunings $\Delta_k = \omega_{k} - \omega^L_k $. 
To achieve TMS, {the cavity detunings must be equal and opposite}: $\Delta_- = - \Delta_+ \approx \omega_m$. 
Transforming to a rotating frame at the pump frequencies and neglecting nonresonant terms, the linearized Hamiltonian in the sideband-resolved regime ($\omega_m \gg \kappa_k$) takes the form

\begin{align}\label{linearized_Hamiltonian}
	H_{OM} = &\Delta_+ a_+^\dagger a_+ + \Delta_- a_-^\dagger a_- + \omega_m b^\dagger b \\
	&+ G_+ (a_+^\dagger b^\dagger + a_+b) + G_- (a_- b^\dagger + a_-^\dagger b)  \nonumber
\end{align}

where the effective optomechanical coupling rates are $G_k = g_k \alpha_k$, with $\alpha_k =- i E_k/(\kappa_k - i\Delta_k)$ (taken to be real without loss of generality). The stability of the system is analyzed in Appendix~\ref{Stability_System} using the Routh–Hurwitz criterion.
The blue-detuned drive ($+$) induces optomechanical parametric amplification and generates entanglement between the cavity and mechanical modes, while the red-detuned drive ($-$) enables state transfer between them. Such a configuration can be realized experimentally, for example, by coupling optical cavities to microwave superconducting circuits \cite{Teufel_om_experiment}, or by employing high-$Q$ mechanical resonators embedded in optical cavity setups \cite{double_cav_alternate}.

\subsection{Modes of Output}

The dynamics of the cavity fields are obtained from the quantum Langevin equations (see Appendix~\ref{Langevin_EOM}). To generate hybrid Bogoliubov modes, the cavity detunings must be appropriately matched. In particular, frequency locking requires $\Delta_- = -\Delta_+ = \omega_m$ \cite{vitali_TMS_generator}, which ensures that the relevant output modes are resonant, by locking the frequencies.

Within the input–output formalism for open quantum systems, the output field operators are related to the intracavity and input noise operators via
\begin{equation}
	a_k^{\mathrm{out}} = \sqrt{\kappa_k}\, a_k - a_k^{\mathrm{in}},
\end{equation}
where $a_k^{\mathrm{in}}$ and $a_k^{\mathrm{out}}$ denote the input and output fields of cavity $k$. In the high-cooperativity regime, $C_- = 4G_-^2/(\kappa_- \gamma_m) \gg 1$, the output modes evaluated at frequencies $\Delta_\pm$ take the form

\begin{subequations} \label{output_bogoliubov}
	\begin{align}
		a_+^{\mathrm{out}} &= \eta_+ b_{\mathrm{in}}^\dagger + \sinh r\, {a_-^{\mathrm{in}}}^\dagger + \cosh r\, a_+^{\mathrm{in}}, \\
		a_-^{\mathrm{out}} &= \eta_- b_{\mathrm{in}} - \sinh r\, {a_+^{\mathrm{in}}}^\dagger - \cosh r\, a_-^{\mathrm{in}} 
	\end{align}
\end{subequations}

where 
$\eta_+ = \frac{ -2i \sqrt{\gamma_m \kappa_+   }  G_+ \kappa_-  }{ G_-^2 \kappa_+ - G_+^2 \kappa_- }$ 
,
$\eta_- = \frac{ -2i \sqrt{\gamma_m \kappa_-} G_- \kappa_+}{ G_-^2 \kappa_+ - G_+^2 \kappa_- }$ 
and 

\begin{align}\label{squeezing_factor}
	\sinh r &= \frac{2 G_- G_+ \sqrt{\kappa_- \kappa_+}}{\kappa_+ G_-^2 - \kappa_- G_+^2}, \
	\cosh r = \frac{G_-^2 \kappa_+ + G_+^2 \kappa_-}{G_-^2 \kappa_+ - G_+^2 \kappa_-}
\end{align}


where $r$ is the squeezing factor, highlighting its dependence solely on the effective optomechanical couplings. The last two terms of each equation in Eq. \eqref{output_bogoliubov} represent TMSV Bogoliubov modes, and the first element is the thermal noise contributed by the reservoir of the mechanical oscillator.

We focus on the steady-state regime and consider filtered output modes centered around the cavity resonances. Independent output modes can be defined by temporal filtering,

\begin{equation}
	f_k^{\mathrm{out}}(t) = \int_{-\infty}^{t} h_k(t - t') \, a_k^{\mathrm{out}}(t') \, \mathrm{d}t',
\end{equation}

where $h_k(t)$ are causal filter functions  satisfying the normalization condition $\int |h_k(t)|^2 \mathrm{d}t = 1$. We adopt exponentially decaying filters of the form

\begin{equation} \label{filterfunction}
	h_k(t) = \sqrt{\frac{2}{\tau}} \, \Theta(t) \, e^{-(1/\tau + i \Omega_k)t},
\end{equation}
where $\Theta(t)$ is the Heaviside step function, $1/\tau$ defines the bandwidth (taken identical for both modes), and $\Omega_k$ is the central frequency measured relative to the corresponding drive.	
Since the output modes are expected to be entangled around $\Delta_\pm$, we fix the central frequencies around $\Omega_k \approx \Delta_k$, while testing squeezing and nonlocality of the output signals. In the following section, we primarily test how resilient the correlation remains with filter parameters, and furthermore with the parameters of the cavities.


\begin{figure*}
	\includegraphics[width= 0.9 \linewidth]{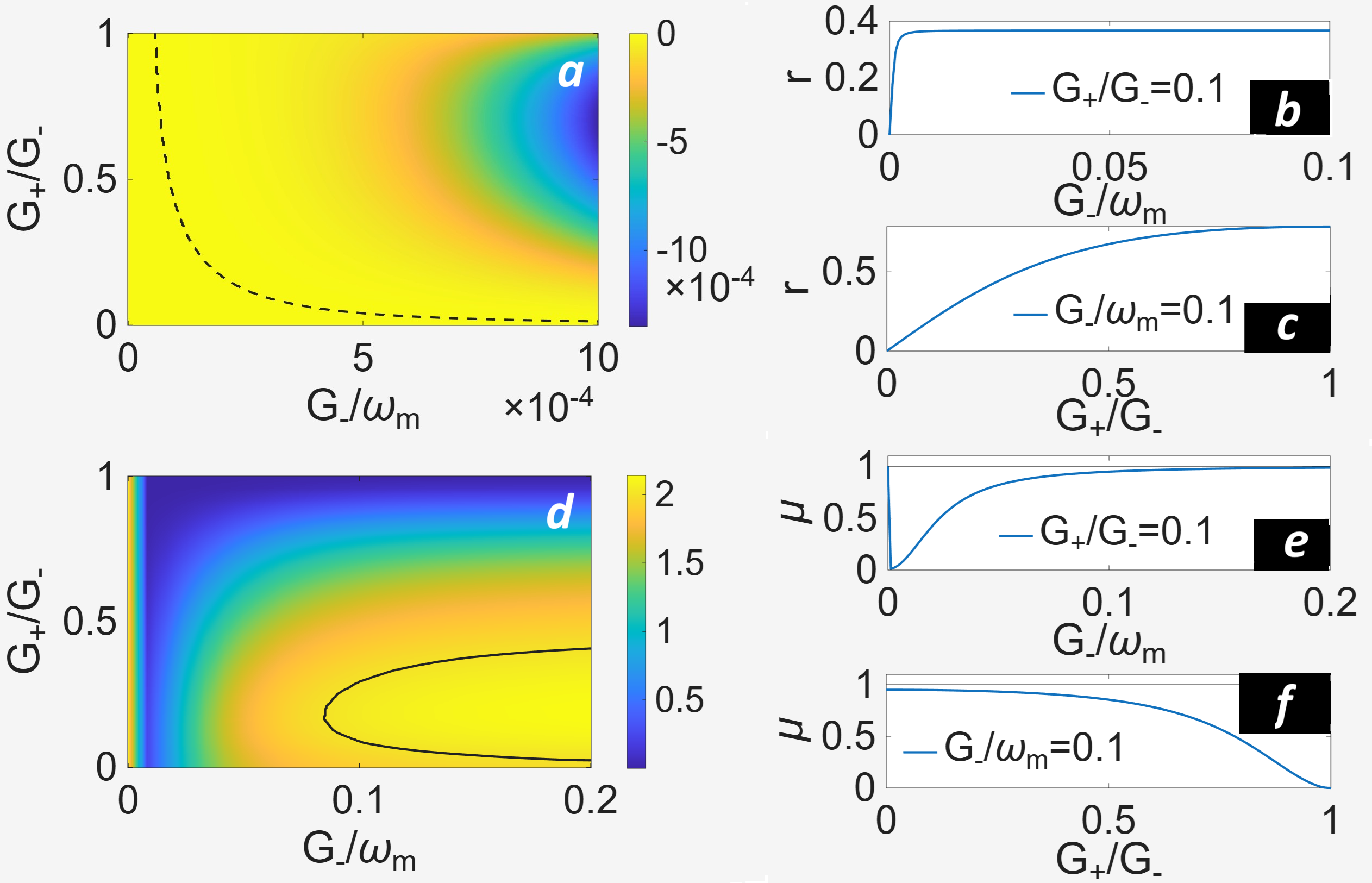}
	\caption{ {(a) $\log[S_q|_{\min}]$ representing TMS as a function of $G_-$ and the ratio $G_+/G_-$. (b) Squeezing factor ($r$) versus $G_-$ for $G_+/G_- = 0.1$.  (c) $r$ versus $G_+/G_-$  for $G_-/\omega_m = 0.1$. (d) The maximal value of Bell function ($B_{max}$) as a function of $G_-$ and the ratio $G_+/G_-$. (e) Purity of the state ($\mu$) versus $G_-$ for $G_+/G_- = 0.1$. (f) $\mu$ versus $G_+/G_-$ for $G_-/\omega_m = 0.1$.  } The black-dashed line in (a) represents SQL ($S_q|_{\min} = 1$) and the black-solid line in (d) stands for the nonlocality limit $(B_{max} = 2)$. All other parameters remain the same with Fig. \ref{SQZFilter_nonloc}.
	}\label{SQZnonlocal}
\end{figure*}

\begin{figure*}
	\includegraphics[width= 0.9 \linewidth]{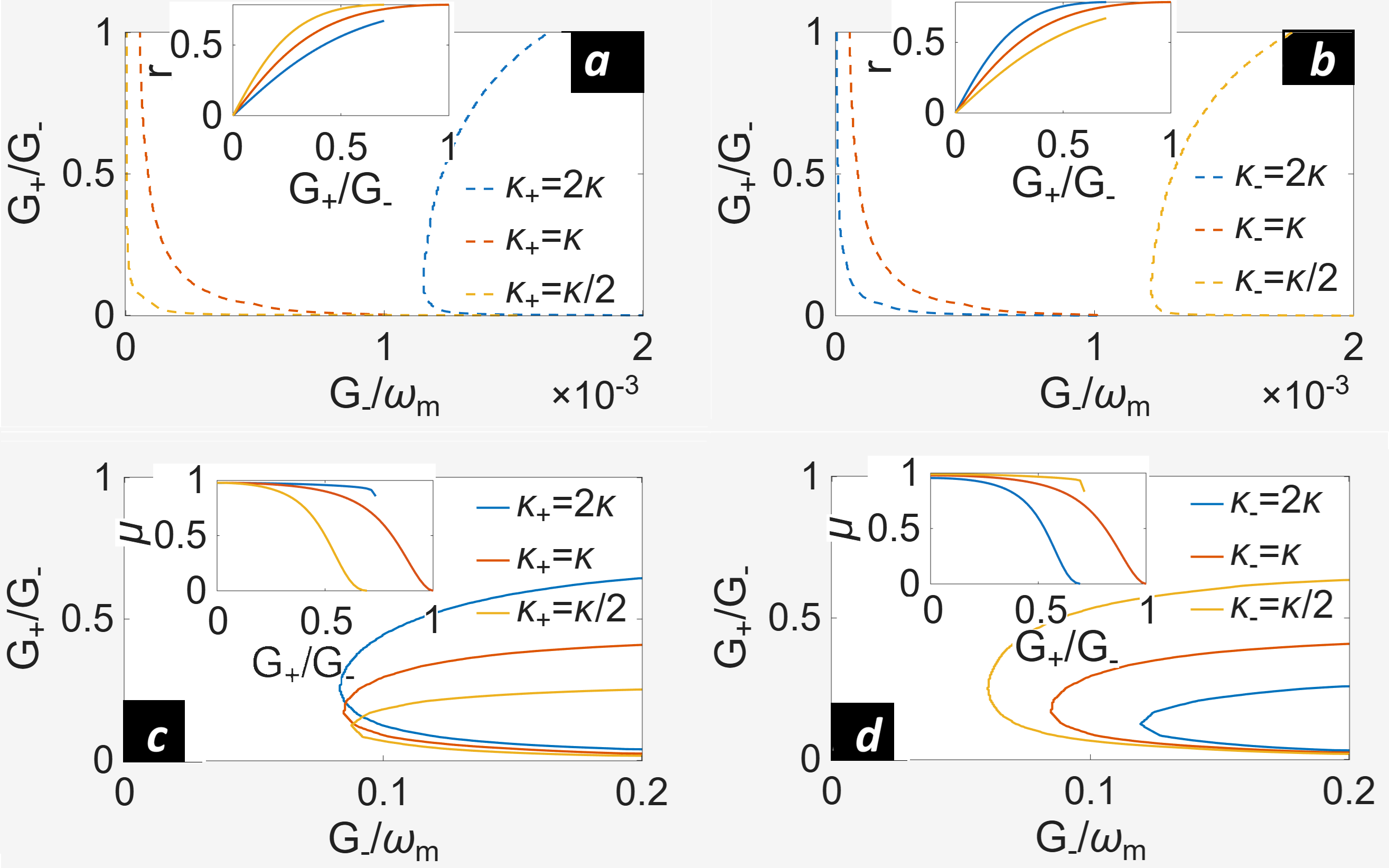}
	\caption{ { (a, b) SQL ($S_q\vert{}_{\min} = 1$) as a function of $G_-$ and $G_+/G_-$ for varying $\kappa_+$ (a) and $\kappa_-$ (b). (c, d) Nonlocality limit ($B_{max} = 2$) as a function of $G_-$ and $G_+/G_-$ for varying $\kappa_+$ (c) and $\kappa_-$ (d). As shown in Fig. \ref{SQZnonlocal}, regions to the right of the dashed boundaries in (a) and (b) exhibit TMS ($S_q\vert{}_{\min} \leq 1$), while regions to the right of the solid boundaries in (c) and (d) exhibit nonlocality ($B_{max} \geq 2$). The insets in (a) and (b) show $r$ versus $G_+/G_-$, and those in (c) and (d) show $\mu$ versus $G_+/G_-$ for $G_- = 0.15\,\omega_m$.} All other parameters are the same as in Fig. \ref{SQZFilter_nonloc}.
	}\label{nonlocal_SQZ_border_k}
\end{figure*}

\begin{figure*}[t!]
	\includegraphics[width= 0.9 \linewidth]{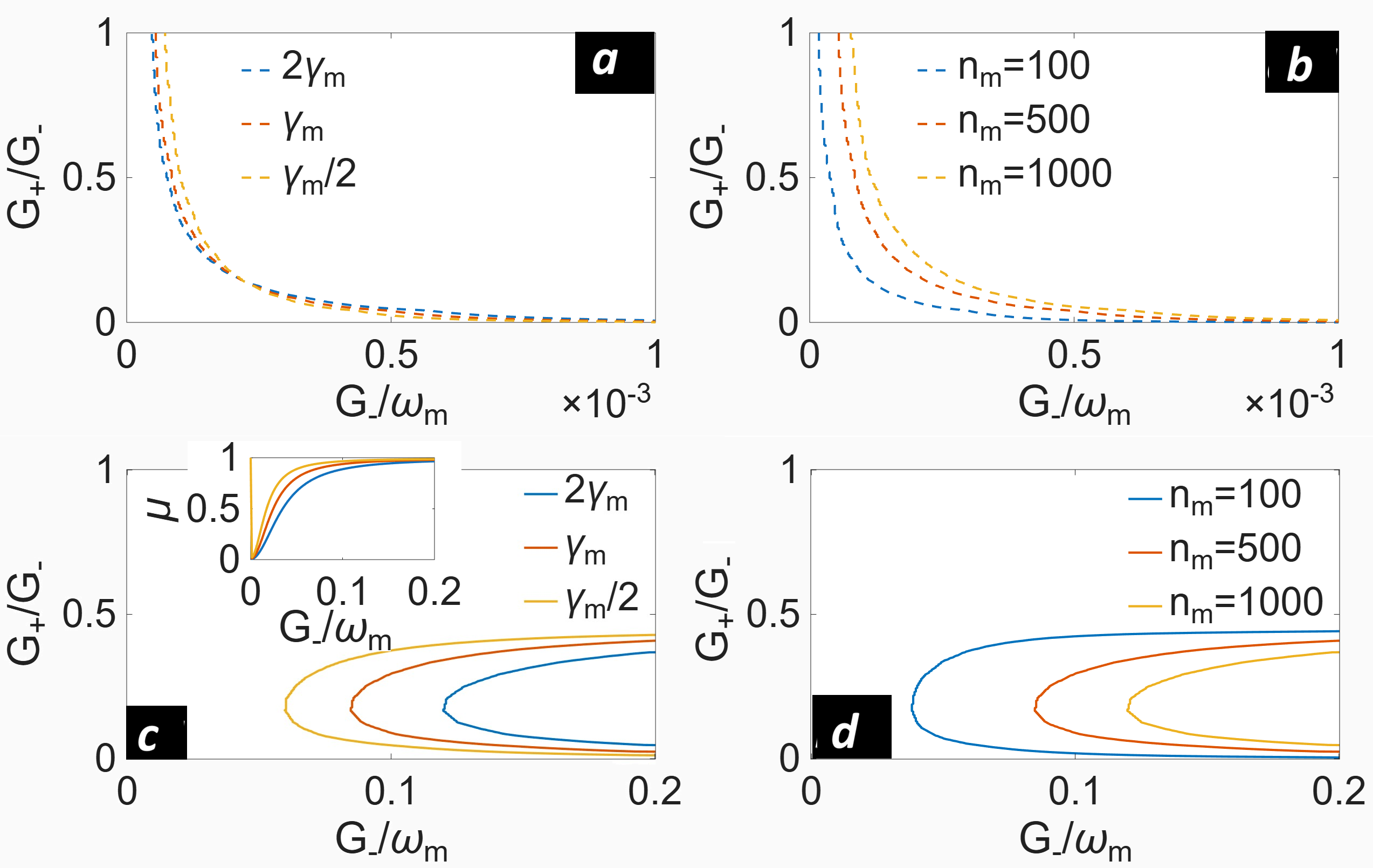}
	\caption{ { (a, b) Standard quantum limit (SQL, $S_q\vert{}_{\min} = 1$) as a function of $G_-$ and the ratio $G_+/G_-$ for varying mechanical oscillator linewidth in (a) and mechanical reservoir temperature in (b). (c, d) Nonlocality limit ($B_{max} = 2$) as a function of $G_-$ and $G_+/G_-$ for varying mechanical oscillator linewidth in (c) and mechanical reservoir temperature in (d). As shown in Fig. \ref{SQZnonlocal}, regions to the right of the dashed boundaries in (a) and (b) exhibit two-mode squeezing ($S_q\vert{}_{\min} \le 1$), while regions to the right of the solid boundaries in (c) and (d) exhibit nonlocality ($B_{max} \ge 2$). The inset in (c) shows $\mu$ versus $G_-$ for $G_+/G_- = 0.2$.} All other parameters are the same as in Fig. \ref{SQZFilter_nonloc}.
	}\label{nonlocal_border_gamma}
\end{figure*}

\section{Results: CHARACTERIZATION of Two mode Squeezing} \label{Results}

\subsection{Impact of Filter frequency and bandwidth}

The entanglement between the two filtered output modes is maximized when the filters select narrowband components centered at the red- and blue-detuned sidebands. This corresponds to choosing the filter frequencies to match the cavity detunings, i.e., $\Omega_- = \Delta_- = -\Delta_+ = -\Omega_+ = \omega_m$. {Here,} we illustrate how the two-mode squeezing (TMS) of the filtered outputs depends on the filter central frequencies and linewidths.	
As a quantitative measure of entanglement, we consider the fluctuations of an optimized hybrid quadrature, as $
S_q(X^{(\phi_+\phi_-)}_{(\mu_+\mu_-)})|_{min} = \frac{1}{2} \langle \{X^{(\phi_+\phi_-)}_{(\mu_+\mu_-)}, X^{(\phi_+\phi_-)}_{(\mu_+\mu_-)}\} \rangle_{min},$
where $X^{(\phi_+\phi_-)}_{(\mu_+\mu_-)}$ denotes a weighted hybrid quadrature, and $\mu_+$ and $\mu_-$ are  weight factors and $\phi_+$ and $\phi_-$ are {arbitrary} phase angles. A detailed definition of the hybrid quadrature and the procedure for obtaining the optimally squeezed quadrature are provided in Appendix~\ref{Maximally_Optimized_Squeezed_Quadrature}. The standard quantum limit (SQL) \textbf{($S_q|_{\min} = 1$)} sets the boundary between separable and entangled states \cite{Duan_Inseparability_entanglement, Vitali_Zippilli_NJP}.
In our analysis,{ we determine the corresponding optimized TMS in Fig.~\ref{SQZFilter_nonloc}(a), while varying the central filter frequency of the blue-detuned cavity for different linewidths. }
In parallel, we evaluate the maximal Bell parameter $B_{\max}$ [Fig.~\ref{SQZFilter_nonloc}(b)] by exploiting correlations between the filtered modes (see Appendix~\ref{nonlocality_bell} for details). The bound $B_{\max} \leq 2$ defines the regime of local realism, and its violation signals nonlocal correlations. {Even though a lot of effort has been invested to study entanglement between the outputs of such system \cite{Clerk_Fidelity, vitali_TMS_generator, Reservoir_Engineering_two_mode}, none of these works examines Bell-nonlocality tests or provides a comparative analysis between squeezing and CHSH violation. Therefore, in this work, we present Bell's measurement on the filtered output of a double-cavity optomechanical system, highlighting the impact of the discrepancy between the central frequencies of the output mode, and analyzing nonlocality comparatively with TMS. }

{
Consistent with expectations, both TMS and Bell violation are maximized for homodyne-like conditions, i.e., when the filters are symmetric ($\Omega_+ = -\Omega_-$), and decrease rapidly as the asymmetry in the filters arises, i.e. the detection moves toward heterodyne-like configurations. This reduction is more pronounced for narrower filter bandwidths (larger $\epsilon$), in agreement with earlier observations of entanglement in the output spectra of similar double-cavity optomechanical systems \cite{vitali_TMS_generator}.
Physically, a narrow bandwidth suppresses frequency mismatch under symmetric filtering, thereby enhancing quantum correlations and leading to stronger TMS and nonlocality. In contrast, for asymmetric (mismatched) filtering, a broader bandwidth increases the effective spectral overlap, facilitating coherent transfer and partially restoring correlations. Similar behavior has been reported in the context of coherent transfer of TMSV states \cite{mypaper_TMSV_filter, mypaper_TMSV_filter_STD}. Comparing between TMS and nonlocality, it is observed that the TMS below SQL persists much broader range than that of nonlocal Bell's measurement, justifying again that, in case of optomechanical outputs, and overall bipartite Gaussian state, the entanglement is necessary, but not sufficient to observe nonlocality \cite{agasti_nonlocally_CV}.
}

\subsection{TMS vs non-Locality}

Since both TMS and nonlocality are maximized under symmetric filtering, $\Omega_- = -\Omega_+ = \omega_m$, we impose this condition to identify experimentally accessible parameter regimes.
{
Fig.~\ref{SQZnonlocal}(a) shows TMS persists over a much broader range, as the double-cavity optomechanical setup provides an efficient source of entangled states. However, the violation of the CHSH inequality emerges only within a finite region of parameter space, predominantly in the large-cooperativity regime, $C_- = 4G_-^2/(\kappa_- \gamma_m) \gg 1$, shown in Fig.~\ref{SQZnonlocal}(d). To elucidate this interplay, we compare the squeezing parameter $r$ with the purity of the state $\mu = \mathrm{Tr}(\rho^2)$ (see Eq. \eqref{purity}) as a function of the effective optomechanical coupling. As shown in Fig.~\ref{SQZnonlocal}(b), increasing cooperativity initially enhances squeezing before reaching saturation, while the purity undergoes a transient reduction followed by recovery (Fig.~\ref{SQZnonlocal}(e)). This behavior explains the simultaneous increase in both TMS and Bell violation observed in Fig.~\ref{SQZnonlocal}(a,d).
A different trend emerges when varying the ratio $G_+/G_-$. Following Eq. \eqref{squeezing_factor}, although the squeezing parameter continues to increase [Fig.~\ref{SQZnonlocal}(c)], the purity of the output state continues decreasing [Fig.~\ref{SQZnonlocal}(f)].This causes the corresponding Bell measurement a nonmonotonic, bell-shaped dependence. CHSH violation is restricted to states with moderate squeezing and sufficiently high purity. In particular, maximal squeezing does not guarantee a violation of the CHSH inequality; instead, nonlocality can be observed for states with lower squeezing. Such behavior, governed by the mixedness of the state, plays a central role in limiting Bell violations, which can also be visualized analytically from Eq. \eqref{nonlocality_formula} given in  Appendix~\ref{nonlocality_bell}. This opens up the possibility to perform Bell's measurement on the scheme presented in \cite{vitali_TMS_generator}, which has widely been accepted for high fidelity non-Gaussian state transfer.
}

These results reinforce the well-known fact that, for bipartite Gaussian CV states, entanglement is necessary but not sufficient for nonlocality \cite{agasti_nonlocally_CV}. Moreover, the Gaussian nature of the states imposes an upper bound on the CHSH parameter ($\lesssim 2.19$), even though the degree of squeezing itself is unbounded.

\subsection{ TMS and non-Locality vs Cavity Finesses}

The most interesting phenomenon is observed when we examine how the boundaries of TMS and nonlocality evolve with the cavity linewidths (or, equivalently, finesse). {
	  Increasing $\kappa_+$, as shown in Fig.~\ref{nonlocal_SQZ_border_k}(a), and decreasing $\kappa_-$ (in Fig.~\ref{nonlocal_SQZ_border_k}(b)) shifts the SQL boundary toward higher cooperativity ($C_-$), thereby reducing the parameter region supporting TMS below SQL. In contrast, the corresponding nonlocality region, surprisingly, expands under the same conditions (Fig.~\ref{nonlocal_SQZ_border_k}(c,d)).
This contrasting behavior can be understood with a comparative analysis of the \sout{\textcolor{red}{the}} degree of squeezing and purity of the states. From Eq.~\eqref{squeezing_factor}, increasing $\kappa_+$ (inset of Fig.~\ref{nonlocal_SQZ_border_k}(a)) and decreasing $\kappa_-$ (inset of Fig.~\ref{nonlocal_SQZ_border_k}(b)) necessarily reduce the achievable squeezing, leading to a contraction of the TMS region. Conversely, at the same time, these changes help preserve the purity of the state (see insets of Fig.~\ref{nonlocal_SQZ_border_k}(c) and (d)), causing increment in Bell's measurement, and thereby enlarging the parameter space over which CHSH violations can occur.  Such non-monotonous behavior between entanglement and non-locality, has even though been observed in case of two-qubit Bell states \cite{inverse_nonlocality_entanglement}, witnessing such event for Gaussian states is a unique achievement. Especially for TMS states, it can open up key practical implications for quantum technology, e.g. optimizing TMS for CV-QKD, not by pushing $r$ too high that reduces nonlocality. }

Note that, despite the expansion of nonlocality region while reducing TMS, the nonlocality region remains strictly contained within the entangled region, reaffirming that entanglement extends beyond the subset of states that exhibit Bell nonlocality.


\subsection{ TMS and non-Locality vs Q factor and temperature }

{
We further investigate how the squeezing and entanglement boundaries evolve with the mechanical quality factor and the temperature of the mechanical reservoir. As shown in Fig.~\ref{nonlocal_border_gamma}(a), increasing the mechanical dissipation rate $\gamma_m$ has only a weak effect on TMS, and the SQL boundary hardly changes. This behavior is consistent with Eq.~\eqref{squeezing_factor}, where the squeezing parameter $r$ does not depend on $\gamma_m$. The small residual shifts in the boundary primarily originate from the coefficients $\eta_k$ in Eq.~\eqref{output_bogoliubov}.
In contrast in Fig.~\ref{nonlocal_border_gamma}(c), the nonlocality boundary decreases rapidly with increasing mechanical linewidth. This trend can be understood in terms of the purity of the state plotted in the inset of Fig.~\ref{nonlocal_border_gamma}(c), which degrades as mechanical dissipation increases, leading to more mixed output states and consequently weaker Bell violations.
}

{
Finally, increasing the thermal occupation of the mechanical reservoir further degrades quantum correlations. The additional thermal noise suppresses both the TMS in Fig.~\ref{nonlocal_border_gamma}(b) and the Bell measurement in Fig.~\ref{nonlocal_border_gamma}(d), thereby reducing the extent of the SQL boundary as well as the region of nonlocality.
}


\section{CONCLUSION}

We have demonstrated how two-mode-squeezed (TMS) output fields can be generated in a double-cavity optomechanical system. Such hybrid quantum interfaces can be prepared experimentally by integrating optical cavities with microwave superconducting circuits \cite{Teufel_om_experiment}, or by employing high-$Q$ mechanical membranes in double-cavity configurations \cite{double_cav_alternate}. This approach provides an alternative route to avoid non-Gaussianity while generating TMS beyond conventional schemes based on parametric down-conversion in Kerr media or four-wave mixing in atomic vapors \cite{squeeze_light_sourse, TMS_generation, Squeezing_generation}, and is well suited for applications in precision measurements \cite{my_BAE} and {high fidelity state transfer for CV teleportation \cite{Clerk_Fidelity}. This, in fact, gives us a unique opportunity to test CHSH violation on the scheme presented in \cite{vitali_TMS_generator}, which offers a promising platform for non-Gaussian state transfer.}
{Moreover, the generation of strongly entangled TMS states using optomechanical systems, and implementing a protocol to perform Bell's measurement on such CV Gaussian states are the novel aspects of this work, remains a broad interest as it avoids non-Gaussianity. In particular, the investigation of Bell nonlocality in such platforms is relevant for constraints imposed by correlated measurements and foundational tests of quantum mechanics \cite{Proposal_Bell_test_Electromechanics, Proposal_Bell_test_optomechanics}.}

By exploiting correlations between spectrally matched output modes, we have investigated the violation of the CHSH Bell inequality. Our results show that the parameter region exhibiting Bell nonlocality is significantly smaller than the region where squeezing falls below the SQL. In particular, maximal squeezing does not necessarily imply Bell violation; instead, nonlocality can arise for states with comparatively lower squeezing. 
We further find that the boundaries of TMS and nonlocality respond differently to variations in system parameters. Notably, changing the cavity linewidths (finesse) causes the TMS and nonlocality regions to shift in opposite directions, a behavior that can be understood in terms of the purity of the state. Additionally, increasing the mechanical dissipation rate reduces the extent of Bell violation, while leaving the TMS region largely unaffected. In contrast, thermal noise in the mechanical reservoir suppresses both squeezing and nonlocality by degrading the underlying quantum correlations. {
	The contrasting behavior between entanglement and Bell's nonlocality is a unique feature discovered here. As entanglement is the cause of nonlocality, a general expectation is monotonous behavior between them. The inverse relation between the entanglement and nonlocality can occasionally be discovered, e.g. in case of spin qubit systems in \cite{inverse_nonlocality_entanglement}. We have experienced it here for the first time in case of CV Gaussian bipartite states, which is a unique finding of the article. In all cases, we rationalize this behavior by analyzing the output-state purity.}

\begin{acknowledgments}
	
	The author is grateful to A. Shukla for his insightful comments while developing the idea and validating the work.
	
\end{acknowledgments}

\section*{funding}
	 The work has been supported by the European Commission, MSCA GA no 101065991 (SingletSQL).

\section*{ Author's contributions}
S.A. conceptualized the idea, developed the methodology, investigated and wrote the manuscript. 

\section*{Availability of Data and Materials}
No data files were generated or analyzed during this study. MATLAB code developed and used for this analysis is available from the corresponding author upon request.

\section*{ Declarations }	

\subsection*{ Conflict of interest/Competing interests}

There is no conflict of interests/competing interests in this paper

\subsection*{ Ethics approval and consent to participate}	

This work has been approved by the Internal Review Board of Hasselt University.

\subsection*{ Disclosures}

The authors declare that there are no conflicts of interest related to this article.

\subsection*{ Consent for publication}

All authors have approved the publication. This research did not involve any human,
animal or other participants.

\subsection*{ Competing interests }
The authors declare no competing interests.

\appendix

\section{Equation of Motion and Correlation Matrix }\label{Correlation_Matrix_Elements}

\subsection{Equation of Motion}\label{Langevin_EOM}

The Heisenberg-Langevin equation \cite{Agasti_OQS} of motion of the system is given by

\begin{equation}
	\dot{\mathbf{u} }  = A \mathbf{u} +\mathbf{u}^{in}  ,
\end{equation}

where	

\begin{align}\label{A_mat}
	A = \left(
	\begin{array}{cccccc}
		-\gamma_m & 0 & 0 & G_+ & 0 & -G_- \\
		0 & -\gamma_m & G_+ & 0 & G_- & 0 \\
		0 & G_+ & -\kappa_+ & 0 & 0 & 0 \\
		G_+ & 0 & 0 & -\kappa_+ & 0 & 0 \\
		0 & -G_- & 0 & 0 & -\kappa_- & 0 \\
		G_- & 0 & 0 & 0 & 0 & -\kappa_- \\
	\end{array}
	\right),
\end{align}

where $\mathbf{u}=[X_+,Y_+,X_-,Y_-,x_m,p_m]^\textbf{T}$ where $X_k = (a_k + a_k^\dagger)/\sqrt{2}, Y_k = -i(a_k - a_k^\dagger)/\sqrt{2}$ are the amplitude and phase quadratures of cavity $k$ and $x_m = (b + b^\dagger)/\sqrt{2}, p_m = -i(b - b^\dagger)/\sqrt{2}$ are for mechanical oscillator, respectively. {Similarly},  $\mathbf{u}^{in}=[\sqrt{2\kappa_+}X_{+}^{in}, \sqrt{2\kappa_+}Y_+^{in}, \sqrt{2\kappa_-}X_-^{in}, \sqrt{2\kappa_-}Y_-^{in}, \sqrt{2\gamma_m}x_m^{in},\sqrt{2\gamma_m}p_m^{in}]^\text{T}$ where $X_{k}^{in} = (a_k^{in} + {a^{in}_k}^\dagger)/\sqrt{2}, Y_{k}^{in} = -i(a_k^{in} - {a_k^{in}}^\dagger)/\sqrt{2}$ and $x^{in}_m = (b^{in} + { b^{in}}^\dagger)/\sqrt{2}, p^{in}_m = -i(b^{in} - {b^{in}}^\dagger)/\sqrt{2}$ are amplitude and phase quadratures and $ a_k^{in}({a_k^{in}}^\dagger),  b^{in}({b^{in}}^\dagger)$ are the {annihilation} (creation) field operators of the inputs of cavity $k$ and mechanical oscillator, respectively. The thermal occupation of the mechanical {reservoir} is $\langle\ b^\dagger b \rangle= n_{m} = (e^{\hbar \omega_m/k_BT}-1)^{-1}$. The correlation function of the input noise to the cavity is given by

\begin{equation}
	\langle  a_k^{in}(t)  {a_k^{in}}^\dagger (t') \rangle = [N(\omega_{k}) +1] \delta(t-t'),
\end{equation} 

where $N(\omega_{k}) = (e^{\hbar \omega_{k}/k_BT}-1)^{-1}$ are the mean thermal occupation numbers of the cavity reservoirs. As the optical frequency is very high $\hbar \omega_{k}/k_BT >>1$, the thermal bath {moreover} behaves as a vacuum $(N(\omega_c) \approx 0)$.

\begin{figure}
	\includegraphics[width= 1 \linewidth]{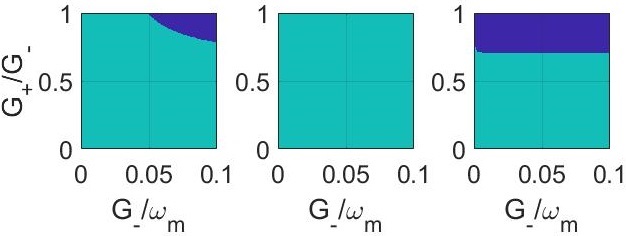}
	\caption{ Stability of the system for $\kappa_+ = 2\kappa$ at left, $\kappa_+ = \kappa$ at center and $\kappa_+ = \kappa/2 $ at right. Green is the stable region, and the blue region is unstable. All other parameters remain the same with Fig. \ref{SQZFilter_nonloc}
	}\label{stability}
\end{figure}

\subsection{Stability of the System} \label{Stability_System}

The stability of the system can be determined by following Routh-Hurwitz stability criteria, yielding a nontrivial constraint which says the real parts of all the eigenvalues of the matrix $A$, given in Eq. \eqref{A_mat} must be negative. In Fig. \ref{stability}, we distinguish the regions where the system can reach a stable state from the region where it does not, and realize that in the case of $\kappa_+ = \kappa_-$, the entire region of $G_+ \leq G_-$ gives a stable solution of the system. However, in the case of  $\kappa_+ \neq \kappa_-$, the region where one can end up obtaining a stable solution reduces significantly, especially for high cooperativity $(C_-)$.

\subsection{Filtered Output}

To determine the output correlation, we need to determine the filtered output quadratures. In the case of TMS output, the filtered quadratures can be determined as

\begin{equation}\label{filtered_quadrature}
	\begin{bmatrix}
		X_k^{f,out} (r,t) \\
		Y_k^{f,out} (-r,t)
	\end{bmatrix} = \int_{-\infty}^{t} \mathrm{d}t' \, T_k(t-t') \begin{bmatrix}
		X_k^{out} (r,t) \\
		Y_k^{out} (-r,t)
	\end{bmatrix},
\end{equation}

where $X_{k}^{out} = (a_k^{out} + {a^{out}_k}^\dagger)/\sqrt{2}, Y_{k}^{out} = -i(a_k^{out} - {a_k^{out}}^\dagger)/\sqrt{2}$,  $X_{k}^{f,out} = (f_k^{out} + {f^{out}_k}^\dagger)/\sqrt{2}, Y_{k}^{f,out} = -i(f_k^{out} - {f_k^{out}}^\dagger)/\sqrt{2}$, and

\begin{equation}
	T_k(t) = \begin{bmatrix}
		\mathcal{R} (h_k^{f,out} ) & -\mathcal{I} (h_k^{f,out} ) \\
		\mathcal{I} (h_k^{f,out} ) & \mathcal{R} (h_k^{f,out} )
	\end{bmatrix},
\end{equation}

where $\mathcal{R}, \mathcal{I}$ represent the real and imaginary parts of the function, respectively. The spectral frequency distribution of the filter function given in Eq. \eqref{filterfunction} in Fourier space is

\begin{equation}
	\tilde{h}_k(\omega) = \frac{\sqrt{\tau/\pi}}{1-i\tau(\omega-\Omega_k)}
\end{equation}

Following Eq. \eqref{filtered_quadrature}, in the frequency domain, the filtered quadratures are given by

\begin{equation}
	\begin{bmatrix}
		\tilde{X}_k^{f,out} (r,\omega) \\
		\tilde{Y}_k^{f,out} (-r,\omega)
	\end{bmatrix} = \sqrt{2\pi} \, \tilde{T}_k(\omega) \begin{bmatrix}
		X_k^{out} (r,\omega) \\
		Y_k^{out} (-r,\omega)
	\end{bmatrix},
\end{equation}
where $\tilde{X}_k^{f,out}(r,\omega),\tilde{Y}_k^{f,out}(r,\omega),\tilde{T}_k(t)$ is the Fourier transformed version of ${X}_k^{f,out}(r,t),{Y}_k^{f,out}(r,t)$ and ${T}_k(t)$.
The elements of the associated covariance matrix of the outputs are

\begin{equation}\label{correlation_mat}
	V_{ij}^{f,out} 
	(\infty) = \frac{1}{2}  \lim_{t \to \infty} \langle \{u_{i}^{f,out}, u_{i}^{f,out} \} \rangle,
\end{equation}

where $\mathbf{u}^{f,out} =[X_+^{f,out},Y_+^{f,out},X_-^{f,out},Y_-^{f,out}]$ where $X_k^{f,out} = (f_k^{out} + {f_k^{out} }^\dagger)/\sqrt{2}, Y_k^{f,out} = -i(f_k^{out} - {f_k^{out} }^\dagger)/\sqrt{2}$ are the amplitude and phase quadratures of respective cavities.

\section{ Squeezing and non-Locality}

\subsection{Maximally Optimized Two-Mode Squeezed Quadrature} \label{Maximally_Optimized_Squeezed_Quadrature}

The weighted quadrature is

\begin{multline}\label{weighted_quadrature}
	X^{(\phi_+\phi_-)}_{(\mu_+\mu_-)} = \frac{1}{\sqrt{\mu_+^2+\mu_-^2}} \bigg[\mu_+ e^{-i\phi_+} f_+^{out} + \mu_+ e^{i\phi_+} {f_+^{out}}^\dagger \\
	+ \mu_- e^{-i\phi_-} f_-^{out} + \mu_- e^{i\phi_-} {f_-^{out}}^\dagger \bigg]
\end{multline}

where $\mu_+,\mu_- \geq 0$ are the weight parameters introduced as the scaling of two systems, and $\phi_+, \phi_-$ are the arbitrary phase angles of the filtered composite quadrature.
The hybrid quadrature variance, defined by $S_q(X^{(\phi_+\phi_-)}_{(\mu_+\mu_-)}) = \frac{1}{2} \langle \{X^{(\phi_+\phi_-)}_{(\mu_+\mu_-)}, X^{(\phi_+\phi_-)}_{(\mu_+\mu_-)}\} \rangle$ becomes

\begin{widetext}
	\begin{align}
		S_q(X^{(\phi_+\phi_-)}_{(\mu_+\mu_-)})  = \frac{2}{\mu_+^2+\mu_-^2} \bigg[ &\mu_+^2  (\cos^2 (\phi_+) V_{11}^{f,out} + \sin^2 (\phi_+) V_{22}^{f,out} + \sin (2\phi_+) V_{12}^{f,out}) \nonumber\\
		& +  \mu_-^2 (\cos^2 (\phi_-) V_{33}^{f,out} + \sin^2 (\phi_-) V_{44}^{f,out} + \sin (2\phi_-) V_{34}^{f,out})  \nonumber\\
		& + 2\mu_+\mu_- \cos(\phi_+) \cos(\phi_-)  V_{31}^{f,out} + 2\mu_+\mu_- \sin(\phi_+) \sin(\phi_-)  V_{24}^{f,out}  \nonumber\\
		& + 2\mu_+\mu_- \cos(\phi_+) \sin(\phi_-)  V_{41}^{f,out} + 2\mu_+\mu_- \sin(\phi_+) \cos(\phi_-)  V_{23}^{f,out}
		\bigg] 
	\end{align}
\end{widetext}

The maximally squeezed quadrature is optimized by {varying} $\mu_+,\mu_-$, and $\phi_+, \phi_-$ to minimize $S_q$, which is furthermore used in Sec. \ref{Results} as the measure of TMS ($S_q|_{min}$). {Since, the two-party output correlation matrix we obtain here are in standard form,} the measurement of maximally optimized TMS is a direct measurement of the entanglement between specific spectral modes \cite{Vitali_Zippilli_NJP}. The squeezing variance can be used to construct entanglement criteria. The maximal squeezing, i.e., minimal TMS variance $(S_q|_{min})$ going below SQL (which is 1) ensures the state to be entangled \cite{Vitali_Zippilli_NJP, Duan_Inseparability_entanglement}.

\subsection{non-Locality: Maximized Bell's function} \label{nonlocality_bell}

The generalized condition for CHSH nonlocality for a CV bipartite Gaussian system is done using the Banaszek-Wodkiewicz phase-space Wigner representation. In this process, one has to transform the correlation matrix $\textbf{V}^{f,out}$ given in Eq. \eqref{correlation_mat} to a standard form using a local linear unitary Bogoliubov operator, and afterwards expressing
the Bell’s function in terms of Wigner functions through Banaszek-W\'odkiewicz representation \cite{agasti_nonlocally_CV}. We find the maximal value of the Bell's function, through investigating over the infinite range of the phase space, as
well as	

\begin{equation}\label{nonlocality_formula}	B_{max} = \mu \left[1 + \left(\frac{ \sqrt{nm}    }{ \sqrt{ nm } +\tilde{c} }\right)^{\frac{    \sqrt{ nm }   }{ \sqrt{ nm } +2\tilde{c} }} \left( \frac{ \sqrt{ n m } +2\tilde{c} }{  \sqrt{ n m } +\tilde{c} }\right)   \right]
\end{equation}

where 
$n^2 = \det(V_+^{f,out}), m^2 = \det(V_-^{f,out}), c_1c_2 = \det(V_{\pm}^{f,out})$ and $\det(\textbf{V}^{f,out}) = (nm-c_1^2)(nm-c_2^2) $, where  $\textbf{V}^{f,out} = \begin{bmatrix}
	\textbf{V}^{f,out}_+ & \textbf{V}^{f,out}_\pm \\
	\textbf{V}^{f,out'}_\pm &\textbf{V}^{f,out}_-
\end{bmatrix}$
and $\tilde{c} = \max[|c_1|,|c_2|]$.

The quantum mechanical {criterion} of the nonlocal realism of a bipartite measurement are the violation of the condition $|B|_{max}\leq 2$. Larger $|B|_{max}$ indicates non-locality to be stronger. The maximum value of $|B|_{max}$ can reach in case of infinitely entangled pure TMS Gaussian states is 2.19 \cite{agasti_nonlocally_CV}. The first term of Eq. \eqref{nonlocality_formula}, given by the mixedness of the state ($\rho$ is the density matrix of the state)

\begin{equation}\label{purity}
	\mu = \text{Tr}[\rho^2]=  \frac{1}{4 \sqrt{\det(\textbf{V}^{f,out} )} }
\end{equation}

By comparing the resulting nonlocality bound with Simon’s separability criterion \cite{Simon}, one can also demonstrate that entanglement is necessary but not sufficient for Bell nonlocality, whereas Bell nonlocality always implies entanglement within the class of bipartite Gaussian states. This provides a unified analytical framework clarifying the boundary between these fundamental quantum correlations \cite{agasti_nonlocally_CV}.

\nocite{*}

\bibliography{apssamp}
\bibliographystyle{apsrev4-2}

\end{document}